\documentclass[12pt]{jpconf}

\usepackage{graphicx}% Include figure files
\usepackage[breaklinks,colorlinks=true,allcolors=blue]{hyperref}
\usepackage{overpic} 
\usepackage{color}
\usepackage{amssymb}
\usepackage{physics} 
\usepackage{newtxtext,newtxmath}
\usepackage{mathtools}
\usepackage{empheq}
\usepackage{natbib}

\usepackage{booktabs} % addlinespace
\newcommand{\pa}{\partial}
\renewcommand{\d}{{\mathrm{d}}}
 
\renewcommand{\i}{\mathrm{i}} 

\newcommand{\fa}[1]{{#1}} 

\begin{document}

\title[]{Relative velocity between a particle and turbulence}

\author{Yutaro Motoori, Daiki Watanabe \& Susumu Goto}

\address{Graduate School of Engineering Science, the University of Osaka \\ 
1-3 Machikaneyama, Toyonaka, Osaka, 560-8531 Japan}
\ead{y.motoori.es@osaka-u.ac.jp, s.goto.es@osaka-u.ac.jp}
\vspace{10pt}
\begin{indented}
\item[]\today
\end{indented}

\begin{abstract}
The relative velocity between a particle and the surrounding fluid is a key
quantity that determines the various phenomena of particle-laden turbulence,
such as particle clustering and turbulence modulation due to particles. We
theoretically derive the expression for the relative velocity of a particle
laden in turbulence by extending the argument of \citet{Balachandar2009}. This
derivation provides a simplified form of the analytical model of
\citet{Berk2024}, and it makes the underlying physical assumptions explicit.
Using direct numerical simulation data of turbulence in a periodic cube and
turbulent channel flow, we demonstrate the validity of the derived expression,
and show that it also holds for non-spherical particles and droplets. 
\end{abstract}

%~~~~~~~~~~~~~~~~~~~~~~~~~~~~~~~~~~~~~~~~~~~~~~~~~
\section{Introduction}
%~~~~~~~~~~~~~~~~~~~~~~~~~~~~~~~~~~~~~~~~~~~~~~~~~

In the present study, we consider the motion of particles in
turbulence. Particle-laden turbulent flows are relevant to a wide range of
phenomena \citep{Subramaniam2022}, such as the formation of rain droplets in clouds
\citep{Vaillancourt2000, Shaw2003, Grabowski2013, Saito2018}, fuel atomization
and combustion in diesel engines \citep{Guzzella2009}, formation of riverbed
sediments \citep{Seminara2010, Scherer2022}, and dispersion of aerosol droplets
\citep{Rosti2021}. The main interests in these studies are twofold: how
particles are transported by turbulence, and how turbulence is modulated by
particles. 

It is well-known that heavy small particles preferentially concentrate in the
regions with low vorticity and high strain rate \citep{Maxey1987, Squires1990,
Squires1991, Wang1993, Bec2007, Soldati2009, Balachandar2010,
Monchaux2012, Gustavsson2016, Brandt2022, Matsuda2024}. Specifically, when the
particle relaxation time is comparable to the Kolmogorov time of turbulence,
particles form clusters around
the smallest vortices. It is also known that as the particle relaxation time
becomes longer, the cluster size also becomes larger \citep{Yoshimoto2007,
Ireland2016, Petersen2019, Berk2020, WangX2020, Oka2021, Oka2021a, Jie2022,
Motoori2022, Motoori2023}. We can explain this multiscale nature of particle
clustering in terms of the presence of multiscale vortices with different 
turnover times in turbulence. 
Our previous studies demonstrated that, using direct numerical simulations (DNS)
of pointwise particles in turbulence in a periodic cube \citep{Yoshimoto2007,
Oka2021, Oka2021a} and turbulent channel flow \citep{Motoori2022, Motoori2023},
heavy particles are swept out from the center of vortices whose turnover time is
comparable to the particle relaxation time, while light particles tend to
accumulate in the vortex center. It is therefore the velocity difference of a
particle and the surrounding fluid that causes these clusterings. \citet{Oka2021a} also
theoretically discussed the scale-dependent clustering rate by evaluating the
divergence of particle velocity field on the basis of the argument by
\citet{Maxey1987}. 

It is also well-known that when particles are added to turbulence, even in a
small amount, the turbulence can be significantly modulated \citep{Tsuji1982,
Tsuji1984, Gore1989, Elghobashi1994, Kajishima2001, Uhlmann2008,
Balachandar2010, Brandt2022, Cannon2024, Chiarini2024, Chiarini2025}. In the
case with negligible gravity, the turbulent kinetic energy tends to be
attenuated \citep{Yu2021, Motoori2025e}. Recent numerical studies
\citep{Oka2022, Balachandar2024, Motoori2025a} showed that the attenuation rate
of turbulent energy is larger when the particle diameter $D$ is smaller (under a
constant volume fraction of particles) or the particle relaxation time is
longer. We can explain this tendency in terms of the additional energy
dissipation rate $\epsilon_p$ around particles. This additional dissipation
bypasses the energy cascade, thereby reducing the energy flux in the inertial
range \citep{Oka2022}. Since the additional dissipation rate can be estimated as
$\epsilon_p \propto \Delta {u}^3 / D$, particles with smaller diameters $D$ or
longer relaxation times (and therefore larger relative velocity $\Delta {u}$)
attenuate turbulence more effectively. Thus, the relative velocity, which
determines the magnitude of $\epsilon_p$, is important for turbulence modulation
due to particles. 

Therefore, the relative velocity plays key roles in particle clustering in
turbulence and turbulence modulation due to particles. In
the present study, we derive a theoretical expression for the relative velocity
by extending the argument by \citet{Balachandar2009}. This
derivation simplifies the model of \citet{Berk2024}, and makes
the physical assumptions explicit. We then verify the derived expression
using direct numerical simulation data of turbulence in a periodic cube
\citep{Oka2022} and turbulent channel flow \citep{Motoori2025a} laden with
spherical finite-size particles. Moreover, using DNS results of turbulence laden
with spheroidal particles \citep{Awai2025} and droplets with a large density
ratio (\ref{sec:W}), we demonstrate that the derived expression is also
applicable to non-spherical particles and droplets.

%~~~~~~~~~~~~~~~~~~~~~~~~~~~~~~~~~~~~~~~~~~~~~~~~~
\section{Theoretical expression of the relative velocity\label{sec:theory}}
%~~~~~~~~~~~~~~~~~~~~~~~~~~~~~~~~~~~~~~~~~~~~~~~~~

%~~~~~~~~~~~~~~~~~~~~~~~~~~~~~~~~~~~~~~~~~~~~~~~~~
\subsection{Assumptions}
%~~~~~~~~~~~~~~~~~~~~~~~~~~~~~~~~~~~~~~~~~~~~~~~~~

We theoretically describe the relative velocity of a particle in turbulence. Let
the particle diameter and mass density be $D$ and $\rho_p$, respectively. Then,
the particle velocity relaxation time is estimated as
\begin{align}
\tau_p = \frac{\rho_p D^2}{18 \rho_f \nu} ,
\label{eq:taup}
\end{align}
where $\rho_f$ and $\nu$ are the fluid mass density and kinematic viscosity,
respectively. The main purpose of the present study is to estimate the relative
velocity $\Delta u$ for a particle with $\tau_p$ in turbulence. To this end, it
is important to identify which scale $\ell$ of flow affects $\Delta u$. In
turbulence, there are eddies with length scales ranging from the Kolmogorov
scale $\eta$ to the integral scale $L$. For scale $\ell$ in the inertial range,
the time scale of eddies is given by $\tau^{(\ell)} = \ell^{2/3}
\epsilon^{-1/3}$. Thus, smaller eddies have shorter time scales of swirling
motion. We identify the size $\ell$ of the eddies which contribute to the
relative velocity under the following two assumptions. (i) Particles cannot
follow fast-changing eddies with length scales smaller than $\ell$; we therefore
neglect these small-scale eddies. (ii) Particles follow slowly-changing eddies
with length scales much larger than $\ell$; accordingly, they have no relative
velocity with respect to these large-scale eddies.

These assumptions imply that the flow scale relevant to the relative velocity
depends on the value of $\tau_p$. When $\tau_\eta < \tau_p < \tau_L$, the two
assumptions mean that eddies at scale $\ell$, which satisfies $\tau^{(\ell)} =
\tau_p$, are relevant to the relative velocity. When $\tau_p < \tau_\eta$, no
eddies have time scales shorter than $\tau_p$. In this case, assumption (ii)
implies that the Kolmogorov scale is relevant to the relative velocity
(i.e.~$\ell=\eta$) because the Kolmogorov-scale eddies have the shortest time
scale. When $\tau_p > \tau_L$, no eddies have time scales longer than $\tau_p$.
In this case, assumption (i) implies that the particle motion is almost
independent of the flow, although the largest-scale eddies weakly affect it
(i.e.~$\ell=L$). In the following subsections, we estimate the relative velocity
for each of these three cases.

%~~~~~~~~~~~~~~~~~~~~~~~~~~~~~~~~~~~~~~~~~~~~~~~~~
\subsection{In the case $\tau_p < \tau_\eta$\label{sec:eta}}
%~~~~~~~~~~~~~~~~~~~~~~~~~~~~~~~~~~~~~~~~~~~~~~~~~

First, we consider the case $\tau_p < \tau_\eta$. Here, we also assume that the
particle diameter $D$ is much smaller than the Kolmogorov length $\eta$ and the
mass density $\rho_p$ of the particle is much larger than that $\rho_f$ of
fluid, because the equation of motion under these assumptions is expressed as
\begin{align}
\frac{\d \vb*{v}_p}{\d t}
=
\frac{1}{\tau_p}
\qty( \vb*{u}(\vb*{x}_p,t)-\vb*{v}_p(t) )
\label{eq:m}
\end{align}
according to \citet{Maxey1983}. In (\ref{eq:m}), $\vb*{v}_p(t)$ is the velocity
of the particle at time $t$, and $\vb*{u}(\vb*{x}_p,t)$ is the fluid velocity at
the particle position $\vb*{x}_p$ and time $t$. We have neglected gravitational effect.

In this case $\tau_p < \tau_\eta$, since there is no fluid motion faster than
$\tau_p$, assumption (i) does not exclude any fluid motion. 
Among the fluid motions that the particle can follow, the
Kolmogorov-scale eddies have the shortest time scale. Therefore, assumption (ii)
implies that the Kolmogorov-scale eddies 
determine the relative velocity. Then, 
we use the one-dimensional model given by
\begin{align}
\frac{\d v_p}{\d t}
=
\frac{1}{\tau_p}
\qty( u(t)-v_p(t) ) 
\label{eq:m-1d} 
\end{align}
similarly to \citet{Balachandar2009}'s argument. However, 
we extend his argument by decomposing the fluid velocity as 
\begin{align} 
u = \overline{u}^{(\eta)} + u^{\prime(\eta)}, 
\label{eq:uslow}
\end{align} 
where $u^{\prime(\eta)}$ is the fluctuating velocity at the Kolmogorov scale and
$\overline{u}^{(\eta)}$ is the more-slowly-changing velocity. This corresponds
to the scale decomposition of the velocity field into the one at scale $\eta$
and those for larger scales than $\eta$. According to assumption (ii), the
particle follows the slowly-changing fluid velocity $\overline{u}^{(\eta)}$, and
therefore we decompose its velocity $v_p$ as
\begin{align}
v_p = \overline{u}^{(\eta)} + v_p^\prime ,
\label{eq:vslow}
\end{align}
where $v_p^\prime$ is the velocity relative to $\overline{u}^{(\eta)}$. By
substituting (\ref{eq:uslow}) and (\ref{eq:vslow}) into (\ref{eq:m-1d}), since
we can neglect $\d \overline{u}^{(\eta)}/\d t$ for the
slowly-changing velocity, we obtain
\begin{align}
\frac{\d v_p^\prime}{\d t}
=
\frac{1}{\tau_p}
\qty( u^{\prime(\eta)}(t)-v_p^\prime(t) ) .
\label{eq:m-eta}
\end{align}
We use (\ref{eq:m-eta}) to estimate the relative velocity in
subsection~\ref{sec:estimate}.

%~~~~~~~~~~~~~~~~~~~~~~~~~~~~~~~~~~~~~~~~~~~~~~~~~
\subsection{In the case $\tau_\eta < \tau_p < \tau_L$\label{sec:ell}}
%~~~~~~~~~~~~~~~~~~~~~~~~~~~~~~~~~~~~~~~~~~~~~~~~~

Next, we consider the case $\tau_\eta < \tau_p < \tau_L$. In this case,
assumption (i) implies that particles cannot follow fluid motions which
change with time scales shorter than $\tau_p$. Then, we can approximate
(\ref{eq:m}) as
\begin{align}
\frac{\d \vb*{v}_p}{\d t}
=
\frac{1}{\tau_p}
\qty( \vb*{u}^{(\ell)}(\vb*{x}_p,t)-\vb*{v}_p(t) ) .
\label{eq:m-cg}
\end{align}
Here, $\vb*{u}^{(\ell)}$ is the fluid velocity coarse-grained at scale $\ell$
with $\ell$ being determined by $\tau^{(\ell)} = \tau_p$. Since $\tau_\eta <
\tau_p = \tau^{(\ell)} < \tau_L$, $\ell$ is in the inertial range, and
therefore, $\tau^{(\ell)} = \ell^{2/3} \epsilon^{-1/3}$. Our previous study
\citep{Oka2021a} demonstrated that (\ref{eq:m-cg}) models the motion of
pointwise particles by numerically integrating (\ref{eq:m-cg}). Accordingly, the
one-dimensional model is expressed as
\begin{align}
\frac{\d v_p}{\d t}
=
\frac{1}{\tau_p}
\qty( u^{(\ell)}(t)-v_p(t) )
\label{eq:1m-cg} .
\end{align}
Here, note that (\ref{eq:m-cg}) and (\ref{eq:1m-cg}) are valid when $D < \ell$
because particles are hardly affected by eddies smaller than $D$.

Then, similarly to the first case, under assumption (ii), we decompose the fluid
velocity as
\begin{align} 
u^{(\ell)} = \overline{u}^{(\ell)} + u^{\prime(\ell)}
\label{eq:uslow-cg}
\end{align} 
and the particle velocity as
\begin{align} 
v_p = \overline{u}^{(\ell)} + v_p^\prime, 
\label{eq:vslow-cg}
\end{align} 
where $u^{\prime(\ell)}$ is the fluid velocity at scale $\ell$,
$\overline{u}^{(\ell)}$ is the more-slowly-changing velocity, and $v_p^\prime$
is the fluctuating velocity of the particle relative to $\overline{u}^{(\ell)}$.
Therefore, we obtain 
\begin{align}
\frac{\d v_p^\prime}{\d t}
=
\frac{1}{\tau_p}
\qty( u^{\prime(\ell)}(t)-v_p^\prime(t) )
\label{eq:m-ell} 
\end{align}
by neglecting $\d \overline{u}^{(\ell)} / \d t$.

%~~~~~~~~~~~~~~~~~~~~~~~~~~~~~~~~~~~~~~~~~~~~~~~~~
\subsection{In the case $\tau_p > \tau_L$\label{sec:L}}
%~~~~~~~~~~~~~~~~~~~~~~~~~~~~~~~~~~~~~~~~~~~~~~~~~

The third case is when $\tau_p > \tau_L$. According to assumption (i), the
particle motion is almost independent of the flow, although it may be weakly
affected by the largest eddies, whose time scale is the largest and
therefore closest to $\tau_p$ among the fluid motions. On the other hand, there
is no fluid motion with time scale longer than $\tau_p$. We therefore retain
only the fluid velocity at the integral scale (i.e.~$u = u^{\prime(L)}$), and
the particle velocity is expressed only by its fluctuating component (i.e.~$v_p
= v_p^\prime$). Then, in this case, we can construct the one-dimensional model
as
\begin{align}
\frac{\d v_p^\prime}{\d t}
=
\frac{1}{\tau_p}
\qty( u^{\prime(L)}(t)-v_p^\prime(t) )
\label{eq:m-L} .
\end{align}
Similarly to the second case, we here assume that $D$ is smaller than $L$,
otherwise particles cannot be affected even by the largest eddies. 

%~~~~~~~~~~~~~~~~~~~~~~~~~~~~~~~~~~~~~~~~~~~~~~~~~
\subsection{Estimation of the relative velocity\label{sec:estimate}}
%~~~~~~~~~~~~~~~~~~~~~~~~~~~~~~~~~~~~~~~~~~~~~~~~~

Looking at the derived one-dimensional models (\ref{eq:m-eta}),
(\ref{eq:m-ell}), and (\ref{eq:m-L}) in
subsections\:\ref{sec:eta}, \ref{sec:ell}, and \ref{sec:L},
respectively, we notice that they are unified as in the form of (\ref{eq:m-ell})
with 
\begin{subequations}
\label{eq:ell}
\begin{empheq}[
left = {{\ell }
= \empheqlbrace \,}, right = {}
]{align}
&
\eta 
&
(&\tau_p < \tau_\eta) ,
\: \label{eq:ell1} \\
%---
&
\tau_p^{3/2} \epsilon^{1/2}
&
(\tau_\eta <\: &\tau_p < \tau_L) , 
\: \label{eq:ell2} \\
%---
&
L
&
(\tau_L <\: &\tau_p) .
\: \label{eq:ell3}
\end{empheq}
\end{subequations}
Then, if we assume $u^{\prime(\ell)}$ and $v_p^\prime$ oscillate with the time-scale of 
\begin{subequations}
\label{eq:tau}
\begin{empheq}[
left = {{\tau^{(\ell)} }
= \empheqlbrace \,}, right = {}
]{align}
&
\tau_\eta 
&
(&\tau_p < \tau_\eta) ,
\: \label{eq:tau1} \\
%---
&
\ell^{2/3} \epsilon^{-1/3}
&
(\tau_\eta <\: &\tau_p < \tau_L) , 
\: \label{eq:tau2} \\
%---
&
\tau_L
&
(\tau_L <\: &\tau_p) ,
\: \label{eq:tau3}
\end{empheq}
\end{subequations}
as
\begin{align}
u^{\prime(\ell)} = \Re \qty[ \widehat{u}^{(\ell)} \: \e^{\i t/\tau^{(\ell)}} ] 
\label{eq:u}
\end{align}
and
\begin{align}
v_p^\prime = \Re \qty[ \widehat{v_p} \: \e^{\i t/\tau^{(\ell)}} ] ,
\label{eq:vp}  
\end{align}
respectively, we obtain
\begin{align}
\widehat{v_p}
=
\frac{\widehat{u}^{(\ell)}}{1+\i \tau_p / \tau^{(\ell)}} .
\label{eq:v0}
\end{align}
Here, $\widehat{u}^{(\ell)}$ and $\widehat{v_p}$ denote the complex amplitudes of
$u^{\prime(\ell)}$ and $v_p^\prime$, respectively. We can therefore estimate
the relative velocity as
\begin{align}
\Delta u
= u - v_p
= u^{\prime(\ell)} - v_p^\prime
= \Re \qty[ \qty( \widehat{u}^{(\ell)} - \widehat{v_p} ) \: \e^{\i t/\tau^{(\ell)}} ] ,
\end{align}
and its amplitude $|\Delta u| = |\widehat{u}^{(\ell)} - \widehat{v_p}
|$ as
\begin{align}
|{\Delta u}|
=
\frac{\tau_p/\tau^{(\ell)}}{\sqrt{1+(\tau_p/\tau^{(\ell)})^2}} \:
|\widehat{u}^{(\ell)}| .
\label{eq:du0}
\end{align}
Using
\begin{subequations}
\label{eq:u-ell}
\begin{empheq}[
left = {{|\widehat{u}^{(\ell)}| }
= \empheqlbrace \,}, right = {}
]{align}
&
u^{\prime(\eta)} %(=\eta^{1/3} \epsilon^{1/3})
&
(&\tau_p < \tau_\eta) ,
\: \label{eq:u-ell1} \\
%---
&
\ell^{1/3} \epsilon^{1/3}
=
(\tau_p/\tau_\eta)^{1/2} u^{\prime(\eta)}
=
(\tau_p/\tau_L)^{1/2} u^{\prime(L)}
&
(\tau_\eta <\: &\tau_p < \tau_L) , 
\: \label{eq:u-ell2} \\
%---
&
u^{\prime(L)} %(=L^{1/3} \epsilon^{1/3})
&
(\tau_L <\: &\tau_p) ,
\: \label{eq:u-ell3}
\end{empheq}
\end{subequations}
we rewrite (\ref{eq:du0}) for each range of $\tau_p$ as
\begin{subequations}
\label{eq:du}
\begin{empheq}[
left = {{|\Delta {u}|}
= \empheqlbrace \,}, right = {}
]{align}
&\frac{\tau_p/\tau_\eta}{\sqrt{1+(\tau_p/\tau_\eta)^2}} \: u^{\prime(\eta)}
&
(&\tau_p < \tau_\eta) ,
\: \label{eq:du1} \\
%---
&\frac{(\tau_p/\tau_\eta)^{\frac{1}{2}}}{\sqrt{2}} \: u^{\prime(\eta)}
=
\frac{(\tau_p/\tau_L)^{\frac{1}{2}}}{\sqrt{2}} \: u^{\prime(L)}
&
(\tau_\eta <\: &\tau_p < \tau_L) , 
\: \label{eq:du2} \\
%---
&\frac{\tau_p/\tau_L}{\sqrt{1+(\tau_p/\tau_L)^2}} \: u^{\prime(L)}
&
(\tau_L <\: &\tau_p) .
\: \label{eq:du3}
\end{empheq}
\end{subequations}

Recall that in the derivation of (\ref{eq:du}), we have assumed that the particle
diameter $D$ is smaller than the eddy scale $\ell$. This prerequisite can be written as 
\begin{subequations}
\label{eq:condition}
\begin{empheq}[
left = { 
D \lesssim \ell
\quad \Leftrightarrow \quad 
D \lesssim
\empheqlbrace 
\, }, right = {}]{align}
& \eta
\hspace{-14mm}
&(&\tau_p < \tau_\eta) ,\: \\
%---
& (\tau_p/\tau_\eta)^{\frac{3}{2}} \: \eta =  (\tau_p/\tau_L)^{\frac{3}{2}} \: L
\hspace{-14mm}
&(\tau_\eta <\: &\tau_p < \tau_L),\: \\
%---
& L
\hspace{-14mm}
&(\tau_L <\: &\tau_p).\: 
%---
\end{empheq}
\end{subequations}
Therefore, the expression (\ref{eq:du}) holds for particles satisfying
(\ref{eq:condition}). Note that $D$ can be larger than $\eta$ for the case
$\tau_p > \tau_\eta$.

Thus, we have derived the expression (\ref{eq:du}) for the relative velocity
which was first derived by \citet{Balachandar2009}. Although the idea of
time-scale matching is similar to his, the present derivation
clarifies the two assumptions (i) and (ii). Under assumption (i),
we coarse-grain the fluid velocity field. This is made by a low-pass filter in
wave-number space, which corresponds to the filter that attenuates
high-frequency fluid fluctuations in \citet{Berk2024}'s model. It is also worth
emphasising that assumption (i) can be formulated by
(\ref{eq:m-cg}) and that assumption (ii) can be implemented
using either (\ref{eq:uslow}) and (\ref{eq:vslow}) for $\tau_p <
\tau_\eta$ or (\ref{eq:uslow-cg}) and (\ref{eq:vslow-cg}) for $\tau_\eta <
\tau_p < \tau_L$. 

Recall that we have developed the above discussion on the basis of the equation
of motion for the particle (\ref{eq:m})
and (\ref{eq:m-cg}), in which the particle motion is described by Stokes drag,
and therefore the particle Reynolds number $Re_p$ is required to be sufficiently
low. In the DNS used for verification in section~\ref{sec:result}, particles
have finite size, with $Re_p \lesssim \mathcal{O}(10^2)$. Therefore, the motion of
these particles is not accurately described by Stokes drag. However,
the essential point of the present discussion is not the detailed form of the
drag law but the time-scale matching between a particle and turbulence.
Therefore, the expression (\ref{eq:du}) of the relative velocity may remain
valid even for $Re_p \gtrsim \mathcal{O}(1)$. In the next section, we verify
(\ref{eq:du}) using DNS data of turbulence in a periodic cube \citep{Oka2022}
and turbulent channel flow \citep{Motoori2025a} laden with finite-size
particles. In addition, we demonstrate its applicability to non-spherical
particles and droplets.

%~~~~~~~~~~~~~~~~~~~~~~~~~~~~~~~~~~~~~~~~~~~~~~~~~
\section{Direct numerical simulations}
%~~~~~~~~~~~~~~~~~~~~~~~~~~~~~~~~~~~~~~~~~~~~~~~~~

We conduct direct numerical simulations of the flow obeying the Navier--Stokes equation,
\begin{align}
\label{eq:NS}
\frac{\pa\vb*{u}}{\pa t}
+
\vb*{u}\cdot\vb*{\nabla}\vb*{u}
=
-
\frac{1}{\rho_f}\vb*{\nabla}p
+
\nu\nabla^2\vb*{u}
+
\vb*{f} ,
\end{align}
and the continuity equation, 
\begin{align}
\label{eq:C}
\vb*{\nabla}
\cdot
\vb*{u}
=
0 ,
\end{align}
where $p(\vb*{x},t)$ and $\vb*{f}(\vb*{x},t)$ are the pressure and the external
force per unit mass, respectively. The equations of motion of a spherical particle are
\begin{align}
\label{eq:p1}
\rho_p V_p
\frac{\d \vb*{v}_p}{\d t}
&=
\oint_{\pa V_p}\: \vb*{\rm T}^{\leftarrow f} \cdot\vb*{n} \:\d S
+
\vb*{F}^{\leftrightarrow p}
\end{align}
and
\begin{align}
\label{eq:p2}
I_p 
\frac{\d \vb*{\omega}_p}{\d t}
&=
\oint_{\pa V_p}\: \vb*{r}\times (\vb*{\rm T}^{\leftarrow f}\cdot\vb*{n})\:\d S
+
\vb*{T}^{\leftrightarrow p} .
\end{align}
Here, $\vb*{v}_p(t)$ and $\vb*{\omega}_p(t)$ are the translational and angular
velocity, respectively, $V_p$ and $I_p$ are the volume and moment of inertia of
a sphere, respectively, $\vb*{\rm T}^{\leftarrow f}$ is the stress tensor for a
Newtonian fluid, and $\vb*{F}^{\leftrightarrow p}$ and $\vb*{T}^{\leftrightarrow
p}$ are the force and torque acting on the particle due to collisions with
other particles or solid walls, respectively. 

For the coupling of fluid and particles, we use the immersed boundary method
\citep{Uhlmann2005, Breugem2012}. Details of the numerical method were described
in \citet{Oka2022} for turbulence in the periodic cube and \citet{Motoori2025a} for turbulent
channel flow. In the following subsections, we show the driving method of
turbulence and particle parameters. 

%~~~~~~~~~~~~~~~~~~~~~~~~~~~~~~~~~~~~~~~~~~~~~~~~~
\subsection{Turbulence in a periodic cube\label{sec:DNS-p}}
%~~~~~~~~~~~~~~~~~~~~~~~~~~~~~~~~~~~~~~~~~~~~~~~~~

The computational domain is a spatially periodic cube with side $2\pi$.
Turbulence is maintained by two types of forcing. One is the time-independent
two-dimensional Taylor--Green force \citep{Goto2017, Goto2024}
\begin{align}
\vb*{f}^{(v)}
=
(-\sin x\:\cos y \:\:\: \cos x\:\sin y \:\:\:  0)^{\rm T} .
\label{eq:fv}
\end{align}
The other is the time-dependent force $\vb*{f}^{(i)}(\vb*{x},t)$
\citep{Lamorgese2005} with the constant energy input rate $P$, whose
Fourier transform is given by
\begin{equation}
\widehat{\vb*{f}^{(i)}}(\vb*{k},t)
=
  \begin{cases*}
    \dfrac{P}{2 E_f(t)} \widehat{\vb*{u}}(\vb*{k},t) & if $\:$ $0<|\vb*{k}|\leq k_f$, \\
    0                 & $\:$ otherwise.
  \end{cases*}
\label{eq:fi}
\end{equation}
Here, $\vb*{k}$ is the wave number, and $\widehat{\vb*{u}}(\vb*{k},t)$ and $E_f$
are the Fourier transforms of $\vb*{u}(\vb*{x},t)$ and kinetic energy contained
in the forcing wave-number range ($0 < |\vb*{k}| \leq k_f$), respectively. The
Taylor-length-based Reynolds number is ${Re}_\lambda = u^\prime \lambda /
\nu = 48$ for the flow forced by $\vb*{f}^{(v)}$ and ${Re}_\lambda = 94$
for $\vb*{f}^{(i)}$, where $u^\prime$ is the root-mean-square of the fluctuating
velocity and $\lambda$ is the Taylor length. 

%~~~~~~~~~~~~~~~~~~~~~~~~~~~~~~~~~~~~~~~~~~~~~~~~~
\subsection{Turbulent channel flow\label{sec:DNS-c}}
%~~~~~~~~~~~~~~~~~~~~~~~~~~~~~~~~~~~~~~~~~~~~~~~~~

The computational domain is set as $4h \times 2h \times 2h$ in the streamwise $x$,
wall-normal $y$, and spanwise $z$ directions, where $h$ is the channel half-width.
We impose the non-slip boundary conditions at $y=0$ and $2h$, and the periodic
boundary conditions in the $x$ and $z$ directions. We drive the flow using the
constant pressure gradient. The friction Reynolds number of the simulated
turbulence is ${Re}_\tau = u_\tau h / \nu = 512$, where $u_\tau$ is the
friction velocity. 

%~~~~~~~~~~~~~~~~~~~~~~~~~~~~~~~~~~~~~~~~~~~~~~~~~
\subsection{Particle parameters}
%~~~~~~~~~~~~~~~~~~~~~~~~~~~~~~~~~~~~~~~~~~~~~~~~~

We numerically track $N_p$ particles with diameter $D$ and mass density
$\rho_p$. 
In the present study, we determine $N_p$ so that the particle
volume fraction is fixed at $\Lambda = 8.2 \times 10^{-3}$, and change $D$ and
$\rho_p$. We show in tables~\ref{table:parameter-p} and \ref{table:parameter-c}
the parameters for turbulence in a periodic cube and turbulent channel flow,
respectively. The particle diameter is larger than the Kolmogorov length (or the
wall-friction length) but smaller than the integral length (or the channel
half-width). The mass density ratio is $\rho_p/\rho_f = 2$, $8$, $32$, $128$,
and $512$. 

In the following, we show the results using the Stokes number $St =
\tau_p/\tau_L$, where $\tau_p$ is the relaxation time defined as
(\ref{eq:taup}), and the non-dimensionalized diameter $D/\eta$ or $D^+$
($=u_\tau D / \nu$). We estimate the large-eddy turnover time $\tau_L$ as
$\tau_L = L/u^\prime$ for turbulence in a periodic cube and $\tau_L = h/u_\tau$
for turbulent channel flow. Particles with $St \lesssim 1$ can follow the
motion of the largest eddies, whereas those with $St \gtrsim 1$ cannot follow
them. The Stokes numbers based on the Kolmogorov time $\tau_\eta$ or the
wall-friction time $\tau^+$, denoted as $St_\eta$ or $St_+$, respectively, are
also listed in tables~\ref{table:parameter-p} and \ref{table:parameter-c}.
These values are larger than unity, indicating that the particles cannot follow
the smallest eddies.

%~~~~~~~~~~~~~~~~~~~~~~~~~~~~~~~~~~~~~~~~~~~~~~~~~
\begin{table}
\centering
	\caption{
		Particle parameters for turbulence in a periodic cube driven by the forces 
 		(a) $\vb*{f}^{(v)}$ and (b) $\vb*{f}^{(i)}$.
    \label{table:parameter-p}}
\vspace{2mm}
\begin{minipage}{0.03\textwidth}
\centering
\begin{tabular}{ccccccccc}
\multicolumn{1}{l}{(\fa{a})  } 
\vspace{48mm}
\end{tabular}
\end{minipage}
\begin{minipage}{0.46\textwidth}
\centering
\begin{tabular}{ccccccccc}
%\multicolumn{6}{l}{(\fa{a})} \\
\addlinespace[2mm]
\multicolumn{1}{c}{$D/\eta$} &
\multicolumn{1}{c}{$D/L$} &
\multicolumn{1}{c}{$\rho_p/\rho_f$} & 
\multicolumn{1}{c}{$St_\eta$} &
\multicolumn{1}{c}{$St$} &
\multicolumn{1}{c}{$N_p$} 
\\ 
\addlinespace[0.6mm]
\hline
\addlinespace[1.0mm]
$7.8$ & $0.17$ & $2$   & $6.5$   & $0.51$ & $512$ \\
$7.8$ & $0.17$ & $8$   & $26$    & $2.0$  & $512$ \\
$7.8$ & $0.17$ & $32$  & $100$   & $8.1$  & $512$ \\
$7.8$ & $0.17$ & $128$ & $410$   & $32$   & $512$ \\
$7.8$ & $0.17$ & $512$ & $1700$  & $130$  & $512$ \\
\addlinespace[1mm]
$16$ & $0.33$ & $2$    & $26$    & $2.0$  & $64$ \\
$16$ & $0.33$ & $8$    & $100$   & $8.1$  & $64$ \\
$16$ & $0.33$ & $32$   & $410$   & $32$   & $64$ \\
$16$ & $0.33$ & $128$  & $1700$  & $130$  & $64$ \\
\vspace{1mm}
\end{tabular}
\end{minipage}
\begin{minipage}{0.03\textwidth}
\centering
\begin{tabular}{ccccccccc}
\multicolumn{1}{l}{(\fa{b})  } 
\vspace{48mm}
\end{tabular}
\end{minipage}
\begin{minipage}{0.46\textwidth}
\centering
\begin{tabular}{cccccccc}
%\multicolumn{5}{c}{(\fa{b})} \\
\addlinespace[2mm]
\multicolumn{1}{c}{$D/\eta$} &
\multicolumn{1}{c}{$D/L$} &
\multicolumn{1}{c}{$\rho_p/\rho_f$} & 
\multicolumn{1}{c}{$St_\eta$} &
\multicolumn{1}{c}{$St$} &
\multicolumn{1}{c}{$N_p$} 
\\ 
\addlinespace[0.6mm]
\hline
\addlinespace[1.0mm]
$8.0$ & $0.15$ & $2$    & $9.0$   & $0.64$ & $512$ \\
$8.0$ & $0.15$ & $8$    & $37$    & $2.6$  & $512$ \\
$8.0$ & $0.15$ & $32$   & $140$   & $10$   & $512$ \\
$8.0$ & $0.15$ & $128$  & $580$   & $41$   & $512$ \\
$8.0$ & $0.15$ & $512$  & $2400$  & $170$  & $512$ \\
\addlinespace[1mm]
$16$ & $0.29$ & $2$    & $37$  & $2.6$  & $64$ \\
$16$ & $0.29$ & $8$    & $140$ & $10$   & $64$ \\
$16$ & $0.29$ & $32$   & $580$ & $41$   & $64$ \\
$16$ & $0.29$ & $128$  & $2400$& $170$  & $64$ \\
\vspace{1mm}
\end{tabular}
\end{minipage}
\end{table}
%~~~~~~~~~~~~~~~~~~~~~~~~~~~~~~~~~~~~~~~~~~~~~~~~~

%~~~~~~~~~~~~~~~~~~~~~~~~~~~~~~~~~~~~~~~~~~~~~~~~~
\begin{table}
\centering
	\caption{
		Particle parameters for turbulent channel flow. The superscript ${\:
		\cdot\:}^+$ denotes quantities normalised by $u_\tau$ and $\nu$. 
    \label{table:parameter-c}}
\vspace{2mm}
\begin{tabular}{cccccccc}
\addlinespace[2mm]
\multicolumn{1}{c}{$D^+$} &
\multicolumn{1}{c}{$D/h$} &
\multicolumn{1}{c}{$\rho_p/\rho_f$} & 
\multicolumn{1}{c}{$St_+$} &
\multicolumn{1}{c}{$St$} &
\multicolumn{1}{c}{$N_p$} 
\\ 
\addlinespace[0.6mm]
\hline
\addlinespace[1.0mm]
$16$ & $0.031$ & $2$    & $28$     & $0.056$ & $8192$ \\
$16$ & $0.031$ & $8$    & $110$    & $0.22 $ & $8192$ \\
$16$ & $0.031$ & $32$   & $460$    & $0.89 $ & $8192$ \\
$16$ & $0.031$ & $128$  & $1800$   & $3.6  $ & $8192$ \\
\addlinespace[1mm]                
$32$ & $0.063$ & $2$    & $110$    & $0.22$  & $1024$ \\
$32$ & $0.063$ & $8$    & $460$    & $0.89$  & $1024$ \\
$32$ & $0.063$ & $32$   & $1800$   & $3.6 $  & $1024$ \\
$32$ & $0.063$ & $128$  & $7300$   & $14  $  & $1024$ \\
\addlinespace[1mm]                
$64$ & $0.13$ & $2$     & $460$    & $0.89$  & $128$  \\
$64$ & $0.13$ & $8$     & $1800$   & $3.6 $  & $128$  \\
$64$ & $0.13$ & $32$    & $7300$   & $14  $  & $128$  \\
$64$ & $0.13$ & $128$   & $29000$  & $57  $  & $128$  \\
\vspace{1mm}
\end{tabular}
\end{table}
%~~~~~~~~~~~~~~~~~~~~~~~~~~~~~~~~~~~~~~~~~~~~~~~~~

%~~~~~~~~~~~~~~~~~~~~~~~~~~~~~~~~~~~~~~~~~~~~~~~~~
\section{Results\label{sec:result}}
%~~~~~~~~~~~~~~~~~~~~~~~~~~~~~~~~~~~~~~~~~~~~~~~~~

Let us verify (\ref{eq:du}) using the DNS data of \citet{Oka2022} and
\citet{Motoori2025a}. More concretely, we focus on (\ref{eq:du2}) and
(\ref{eq:du3}) since the particle relaxation time is larger than the Kolmogorov
time (i.e.~$\tau_p \gg \tau_\eta$) in the present DNS. 

%~~~~~~~~~~~~~~~~~~~~~~~~~~~~~~~~~~~~~~~~~~~~~~~~~
\begin{figure}[!t]
	\centering
	\begin{overpic}[clip,width=1.0\textwidth]{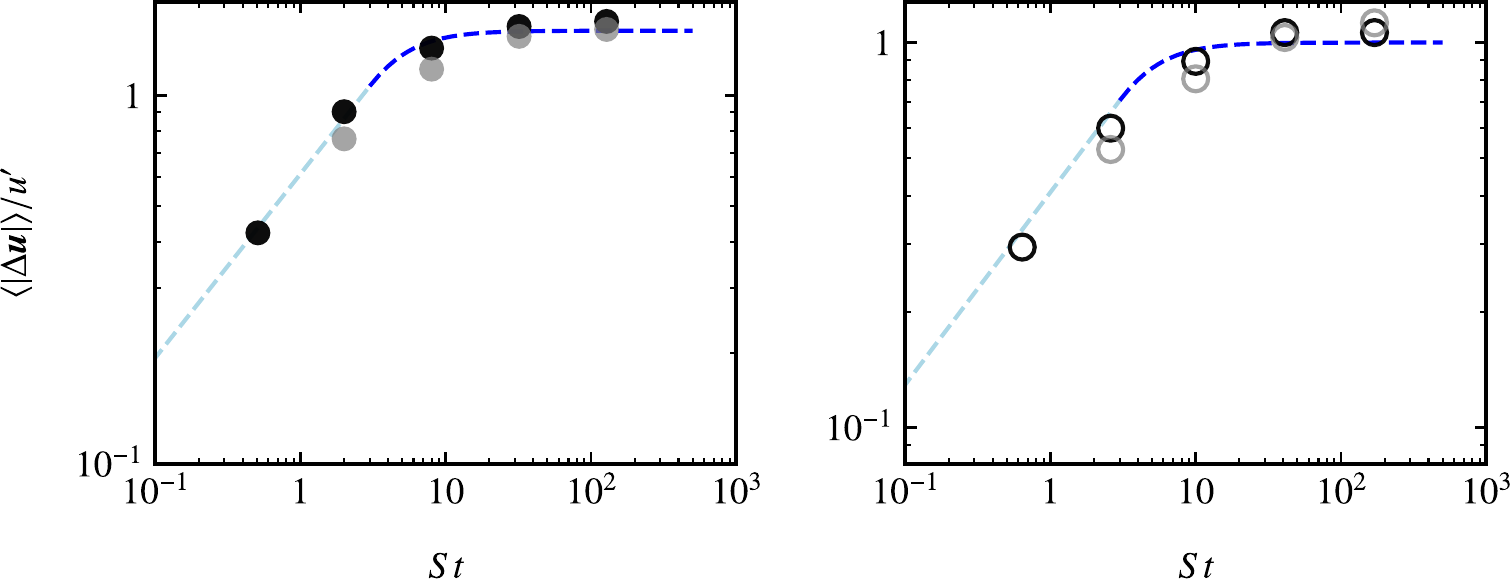}
	\put( 0,37){(\fa{a})}
	\put(51.5,37){(\fa{b})}
	\end{overpic}
	\vspace{-3mm}
	\caption{Average relative velocity $\langle |\Delta \vb*{u}| \rangle$ in the
	turbulence in a periodic cube driven by the forces (\fa{a}) $\vb*{f}^{(v)}$
	and ({\fa{b}}) $\vb*{f}^{(i)}$. The black symbols indicate (a) $D/\eta =
	7.8$ and (b) $8.0$, and the grey ones indicate (a) $D/\eta = 16$ and (b)
	$16$. The light and dark blue dashed lines are (\ref{eq:du2}) and
	(\ref{eq:du3}), respectively, with (a) $u^{\prime(L)} = 1.5u^\prime$ and (b) $u^{\prime(L)} =
	u^\prime$. \label{fig:du-p} }
\end{figure}
%~~~~~~~~~~~~~~~~~~~~~~~~~~~~~~~~~~~~~~~~~~~~~~~~~

%~~~~~~~~~~~~~~~~~~~~~~~~~~~~~~~~~~~~~~~~~~~~~~~~~
\begin{figure}[!t]
	\centering
	\begin{overpic}[clip,width=0.55\textwidth]{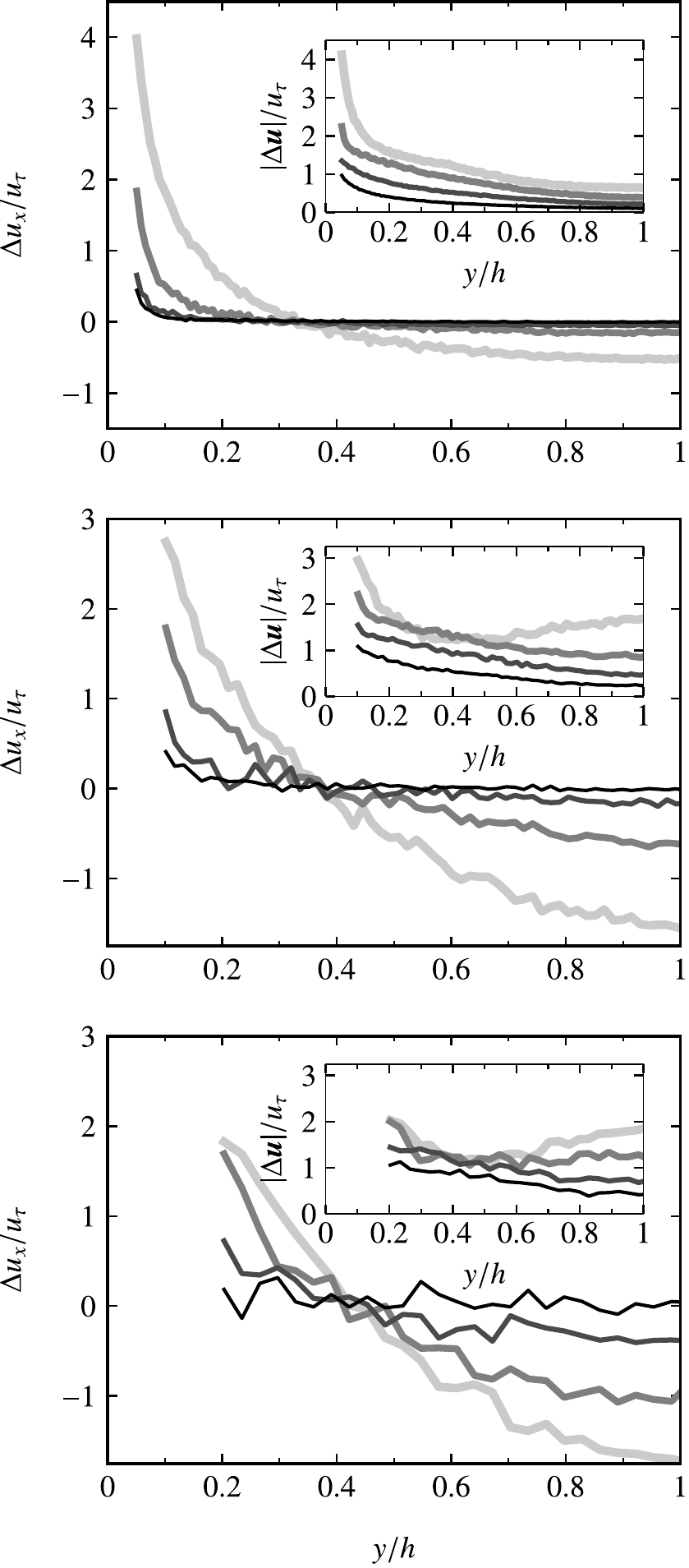}
	\put( 0,99.3){(\fa{a})}
	\put( 0,66.3){(\fa{b})}
	\put( 0,33.3){(\fa{c})}
	\end{overpic}
	\vspace{-3mm}
	\caption{Wall-normal profile of the streamwise component $\Delta u_x$ of the
	mean relative velocity in turbulent channel flow laden with particles of (a)
	$D^+ = 16$, (b) $32$, and (c) $64$. Thicker and lighter lines indicate
	larger $St$: namely, (a) $St=0.056$, $0.22$, $0.89$, and $3.6$, (b) $0.22$,
	$0.89$, $3.6$, and $14$, and (c) $0.89$, $3.6$, $14$, and $57$. The inset
	shows the mean magnitude $|\Delta \vb*{u}|$ of the relative velocity.
	\label{fig:du-c}
	}
\end{figure}
%~~~~~~~~~~~~~~~~~~~~~~~~~~~~~~~~~~~~~~~~~~~~~~~~~

We show in figure~\ref{fig:du-p} the results for turbulence in a periodic cube
driven by (\fa{a}) $\vb*{f}^{(v)}$ and (\fa{b}) $\vb*{f}^{(i)}$. Here, $\langle
|\Delta \vb*{u}| \rangle$ denotes the magnitude of the relative
velocity between a particle and the surrounding fluid, averaged over time and
particles in the statistically steady state. We evaluate the relative velocity
$\Delta \vb*{u} = \vb*{u}_p - \widetilde{ \vb*{u} }^{(p)}$, where
$\widetilde{ \vb*{u} }^{(p)}$ is the fluid velocity around a particle, defined
as the average velocity over the surface of a sphere of diameter $2D$ concentric
with the particle \citep{Kidanemariam2013, Uhlmann2017}. In
figure~\ref{fig:du-p}, as the Stokes number $St$ increases, $\langle |\Delta
\vb*{u}| \rangle / u^\prime$ also increases, and approaches $\mathcal{O}(1)$. We
see this tendency irrespective of the particle diameter, which is indicated by
the color of the symbols. The light and dark dashed blue lines correspond to
(\ref{eq:du2}) and (\ref{eq:du3}), respectively. It is important to observe
that the DNS data points collapse onto these lines. Here, we set
the characteristic velocity $u^{\prime(L)}$ in the order of $u^\prime$
[i.e.~(\fa{a}) $u^{\prime(L)} = 1.5u^\prime$ and (\fa{b}) $u^{\prime(L)} =
u^\prime$]. The larger prefactor of $u^{\prime(L)}$ in (\fa{a}) is due to the
mean flow generated by the forcing, which is absent in (\fa{b}).

Next, we show the results for the turbulent channel flow. In wall-bounded
turbulence, the relative velocity depends on the distance from the wall.
Figure~\ref{fig:du-c} shows the streamwise component of the relative velocity,
averaged over time and particles located at height $y$, for (a) $D^+=16$, (b)
$32$, and (c) $64$. The thicker and lighter lines indicate larger values of the
Stokes number $St$. The insets show the wall-normal profiles of the magnitude of
the relative velocity $|\Delta \vb*{u}|$. Irrespective of the particle diameter,
$|\Delta \vb*{u}|$ becomes larger for larger $St$. We also see that the sign of
the averaged relative velocity changes around $y/h \approx 0.3$ for $St \gtrsim
1$. This indicates that since particles with $St \gtrsim 1$ cannot follow the
swirl of outer-layer-scale vortices (at $0.2$--$0.3h$), particles move slower
than the surrounding fluid in the outer layer; whereas they move faster in the
slower inner layer. Thus, the relative velocity depends on the distance from the
wall, and exhibits qualitatively different behaviour between the outer and inner
layers. 

%~~~~~~~~~~~~~~~~~~~~~~~~~~~~~~~~~~~~~~~~~~~~~~~~~
\begin{figure}[!t]
	\centering
	\begin{overpic}[clip,width=0.7\textwidth]{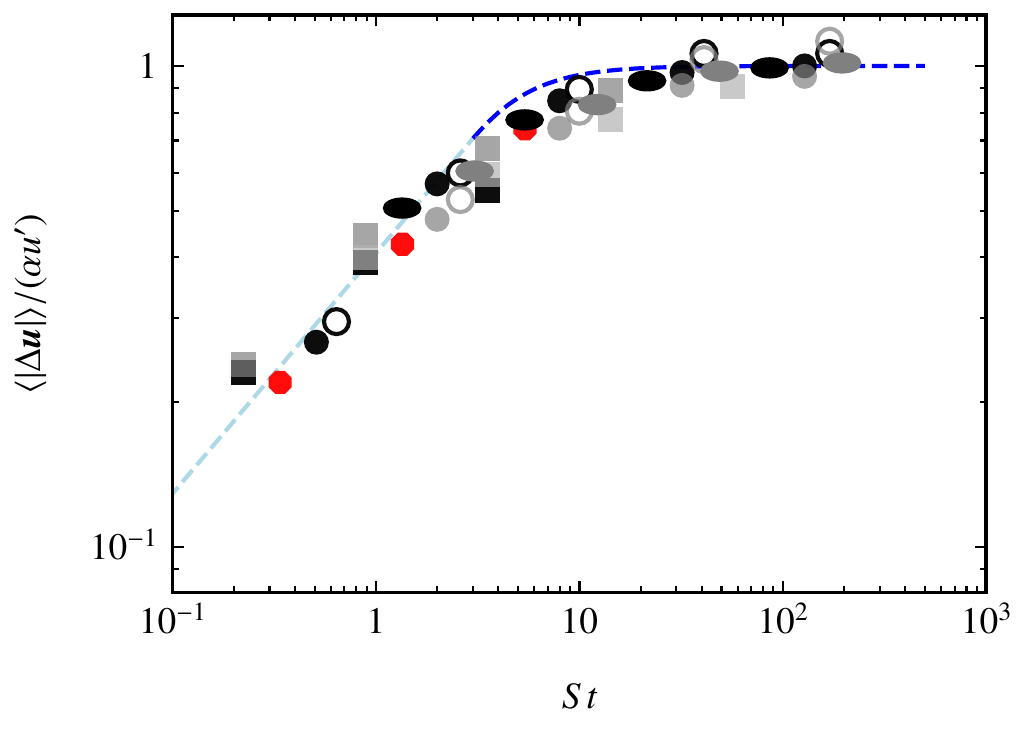}
	\end{overpic}
	\vspace{-3mm}
	\caption{Average relative velocity $\langle |\Delta \vb*{u}| \rangle$ in
	turbulence in a periodic cube (circles, same as in figure~\ref{fig:du-p})
	and in the outer layer of turbulent channel flow (squares). The light and
	dark blue dashed lines correspond to (\ref{eq:du2}) and (\ref{eq:du3}),
	respectively, with $u^{\prime(L)} = \alpha u^\prime$. For turbulence in the
	periodic cube, $\alpha = 1.5$ for $\vb*{f}^{(v)}$; and $\alpha = 1$ for
	$\vb*{f}^{(i)}$. For turbulent channel flow, $\alpha = 1$, and $u^\prime$ is
	the root-mean-square velocity fluctuation in the outer layer. Results for
	prolate spheroidal particles with an aspect ratio of $7$
	\citep{Awai2025} are shown by elliptical symbols (black for equivalent
	diameter $D_*/L = 0.16$, grey for $0.24$), and those for droplets
	(see~\ref{sec:W}) by red symbols.
	\label{fig:du}
	}
\end{figure}
%~~~~~~~~~~~~~~~~~~~~~~~~~~~~~~~~~~~~~~~~~~~~~~~~~

To verify (\ref{eq:du}) for turbulent channel flow, we evaluate the relative
velocity $\langle |\Delta \vb*{u}| \rangle$ averaged in the outer layer ($y/h
\geq 0.3$). We plot in figure~\ref{fig:du} the values of $\langle |\Delta
\vb*{u}| \rangle$ by squares, together with the results for turbulence in the
periodic cube (circles, the same as in figure~\ref{fig:du-p}). For turbulent
channel flow, as $St$ gets larger, the relative velocity approaches the
characteristic velocity $\alpha u^\prime$ of the largest eddies, where $\alpha$
is an order-one coefficient depending on the turbulence field. We can observe
that all data points are distributed around the dashed line indicating
(\ref{eq:du}). This means that in both wall-bounded and unbounded turbulence,
the mean relative velocity is well approximated by (\ref{eq:du}).

Moreover, (\ref{eq:du}) can be used for non-spherical particles;
figure~\ref{fig:du} shows results \citep{Awai2025} for spheroidal particles in
turbulence in the periodic cube driven by the force $\vb*{f}^{(v)}$
(\ref{eq:fv}), which are indicated by elliptical symbols. The color represents
the particle diameter. Although the particles are elongated with an aspect ratio
of $7$, (\ref{eq:du}) gives a good approximation of the relative velocity. Since
(\ref{eq:du}) is derived considering only the translational velocity of
particles, this result implies that rotational motions make only a small
contribution to the relative velocity. This is because the rotational relative
velocity $\Delta u_r$ is determined by eddies of size comparable to the particle
diameter (i.e.~$\Delta u_r \sim D_*^{\frac{1}{3}} \epsilon^{\frac{1}{3}}$), and
it is smaller than the translational one; see \citet{Awai2025} for details of
this argument.

In addition, the red symbols in figure~\ref{fig:du} show the average relative
velocity of droplets in turbulent channel flow. These results are obtained by
DNS for two-phase flow developed by \citet{Watanabe2024} and
\citet{Watanabe2025} (see also \ref{sec:W} for the numerical methods and
parameters). We see that (\ref{eq:du}) also holds for small droplets satisfying
(\ref{eq:condition}). For droplets, the relaxation time (\ref{eq:taup}) is
modified as $\tau_d = C_\mu \rho_d D^2/(18\rho_c\nu_c)$ with
$C_\mu=(3+3\mu_d/\mu_c)/[2+3(\mu_d/\mu_c)]$ \citep{Loth2008}, where the
subscripts $d$ and $c$ denote the dispersed and continuous phases, respectively.
The expression (\ref{eq:du}) can be therefore used for droplets by defining the
Stokes number with this viscosity-ratio-dependent relaxation time. Since the
correction factor $C_\mu$ is close to unity for the present parameters, with a
maximum value of $1.13$ (\ref{sec:W}), we have used the uncorrected $St$ in
figure~\ref{fig:du}.

%~~~~~~~~~~~~~~~~~~~~~~~~~~~~~~~~~~~~~~~~~~~~~~~~~
\section{Conclusions}
%~~~~~~~~~~~~~~~~~~~~~~~~~~~~~~~~~~~~~~~~~~~~~~~~~

In the present study, we have discussed the relative velocity of particles
smaller than the integral length $L$ in turbulence. First, in
section~\ref{sec:theory}, we have derived the theoretical expression
(\ref{eq:du}) of the relative velocity by extending the argument of
\citet{Balachandar2009}. This derivation also provides a simplified form of the
analytical model of \citet{Berk2024}. Then, in section~\ref{sec:result}, we have
verified (\ref{eq:du}) for $\tau_p > \tau_\eta$ using our DNS results of
turbulence in a periodic cube \citep{Oka2022} and turbulent channel flow
\citep{Motoori2025a}. We have also demonstrated that (\ref{eq:du}) is applicable
to non-spherical particles and droplets. As described in the introduction,
expression (\ref{eq:du}) provides a useful basis for modelling particle
clustering and turbulence modulation in particle-laden flows.

%~~~~~~~~~~~~~~~~~~~~~~~~~~~~~~~~~~~~~~~~~~~~~~~~~
\section*{Acknowledgments}
%~~~~~~~~~~~~~~~~~~~~~~~~~~~~~~~~~~~~~~~~~~~~~~~~~

This study was partly supported by the JSPS Grants-in-Aid for Scientific
Research (25K01158 and 26K17306) and Transformative Research Areas (26H00389).
The DNS were conducted by using the computational resources of the
supercomputers Fugaku through the HPCI System Research Projects (hp240278 and
hp260139). The numerical analyses were conducted under the auspices of the NIFS
Collaboration Research Programs (NIFS24KISC007 and NIFS26KISC034). 

\appendix

%~~~~~~~~~~~~~~~~~~~~~~~~~~~~~~~~~~~~~~~~~~~~~~~~~
\section{Numerical method for turbulent channel flow with a droplet
\label{sec:W}}
%~~~~~~~~~~~~~~~~~~~~~~~~~~~~~~~~~~~~~~~~~~~~~~~~~
We conduct DNS of turbulent channel flow with a droplet by solving the
two-phase Navier--Stokes equation. The numerical procedure is based on that of
\citet{Watanabe2024}, which was developed for gas--liquid two-phase flow.
Therefore, to enable the DNS of liquid--liquid two-phase flow, we make two
modifications to the numerical procedure. First, we use the level-set method
\citep{Sussman1994} combined with a global mass preservation method
\citep{yap2006} to capture the motion of the droplet interface. Second, we
incorporate the surface tension force based on the density-scaled continuum
surface force model \citep{Yokoi2014}. On channel walls, we set the contact
angle to $150^\circ$ \citep{sussman2001}.

We evaluate the advection terms in both the Navier--Stokes equation and the
equation for the level-set function using the third-order WENO-P+3 scheme
\citep{Xu2018}. For the time integration of the latter, we use the third-order
Runge--Kutta method \citep{Gottlieb2001}.

We have included in figure~\ref{fig:du} the relative velocity of a droplet in
turbulent channel flow at $Re_\tau = 180$.  The computational domain size and
spatial resolution are the same as those of our DNS for solid particles at
$Re_\tau = 180$ \citep{Motoori2025a}. We fix the initial droplet diameter $D$ at
$0.13$, and change the density ratio $\rho_d/\rho_c = 2$, $8$, and $32$, which
correspond to $St = 0.34$, $1.4$, and $5.4$, respectively. We also set the
viscosity ratio equal to the density ratio, namely, $\mu_d/\mu_c =
\rho_d/\rho_c$ ($= 2$, $8$, and $32$). In addition, we choose the surface
tension $\sigma$ (the Weber number is $We = {\rho_d u_\tau^2 D} / {\sigma} =
0.65$) so that the droplet does not break up.

To accurately evaluate the relative velocity, we eliminate coalescence due to
droplet collisions by introducing a single droplet into a turbulent channel
flow. After the flow becomes a statistically steady state, we take the temporal
average of the relative velocity for a period of $t \gtrsim h/u_\tau$. We
also take the average for $20$ different initial conditions.

\end{document}